\documentclass{article}
\usepackage{spconf, amsmath, graphicx, rotating, adjustbox, booktabs, bm, float, amsfonts}
\usepackage[table]{xcolor}
\usepackage{amssymb}
\usepackage{mathtools}
\usepackage{multirow}

\title{Investigating Self-Supervised Deep Representations for EEG-based Auditory Attention Decoding}
\name{Karan Thakkar, Jiarui Hai, Mounya Elhilali\thanks{This work was supported by ONR N00014-23-1-2050 and N00014-23-1-2086 and NIH U01AG058532}} 
\address{Laboratory for Computational Audio Perception, Johns Hopkins University, USA}
\begin{document}
%
\maketitle
\begin{abstract}
Auditory Attention Decoding (AAD) algorithms play a crucial role in isolating desired sound sources within challenging acoustic environments directly from brain activity. Although recent research has shown promise in AAD using shallow representations such as auditory envelope and spectrogram, there has been limited exploration of deep Self-Supervised (SS) representations on a larger scale. In this study, we undertake a comprehensive investigation into the performance of linear decoders across 12 deep and 2 shallow representations, applied to EEG data from multiple studies spanning 57 subjects and multiple languages. Our experimental results consistently reveal the superiority of deep features for AAD at decoding background speakers, regardless of the datasets and analysis windows. This result indicates possible nonlinear encoding of unattended signals in the brain that are revealed using deep nonlinear features. Additionally, we analyze the impact of different layers of SS representations and window sizes on AAD performance. These findings underscore the potential for enhancing EEG-based AAD systems through the integration of deep feature representations.

\end{abstract}
\begin{keywords}
auditory attention decoding, electroencephalogram (EEG), self-supervised speech representations 
\end{keywords}
\section{Introduction}
\label{sec:intro}


Throughout daily life, we all experience the challenges of following a particular conversation in presence of other competing speakers or noise sources. This challenge is particularly pronounced in individuals with hearing impairment and impedes their ability to interact socially. This process relies on our brain's attentional mechanisms in order to hone in on a speaker of interest and render the rest of the acoustic scene to the background. Auditory attention decoding (AAD) is a general framework developed to determine the sound a listener is attending to based on their brain activity; hence holding the potential to improve hearing aids and neuroprosthetics \cite{Van_2017}. Various methods to collect brain data have shown promise for AAD, though they come with different trade-offs in terms of invasiveness, portability, resolution, and signal quality. Non-invasive electroencephalography (EEG), for instance, relies on scalp electrodes to capture signals \cite{o2015attentional}. It is also the most portable and adaptable allowing the user to potentially go about their daily life; though it comes at a cost of lower spatial resolution and signal quality. On the other hand, magnetoencephalography (MEG) offers higher resolution in mapping brain activity patterns, but demands bulky and expensive equipment for its operation \cite{akram2016robust}. Alternative invasive methods such as electrocorticography (ECoG) implant electrodes directly into the brain, providing unparalleled precision but also exposing the patient to surgical risk.

Across all these techniques, AAD learns a mapping between the complex auditory stimuli entering the ears and the brain activity patterns generated in response \cite{o2015attentional}. In its simplest form, this mapping can be approximated by a linear function (e.g. regression) by  learning correspondence between brain signals and the attended envelope or spectrogram of the foreground speaker \cite{crosse2016multivariate}. These techniques have shown great promise for AAD given their simplicity and minimal training and data needs. In contrast, deep learning excels at learning hierarchical abstractions, allowing it to identify intricate relationships between auditory inputs and neural representations. Recent work has shown promise in applying mappings learned through CNNs or LSTMs \cite{CNN-AAD, accou2023decoding}; though by-and-large, these models are data-and computation-hungry and require extensive tuning. 

Nevertheless, deep learning representations
offer the potential to capture intricate and nuanced connections between auditory stimuli and neural responses, which may elude simpler linear models, particularly in shedding light on how the brain disentangles the representation of foreground (attended) and background (unattended) information. Building on this potential, this study performs a meta-analysis of a range of deep and shallow features, all evaluated within the \emph{same} framework on publicly available datasets, spanning different research subjects and languages. The study addresses the following research questions:  \textbf{1)} How do deep features fair against shallow features in a direct AAD comparison? \textbf{2)} Do abstractions learned through deep features reveal distinctions in how the brain represents foreground and background sensory signals? \textbf{3)} How generalizable are deep features trained on one language to other languages when applied to AAD?

\section{AAD Methodology}
\label{sec:format}

\subsection{General AAD framework}

The general AAD framework learns the mapping between an audio signal $a(t)$ and the brain response $r(t,n)$, where $t$ is the index of time and $n$ represents different neural channels. This mapping is often estimated in two steps (Eq.\ref{eq.AAD}): 

\begin{equation}
    a(t) \xrightarrow{\Gamma(.)} s(t) \xleftrightarrow{\Psi(.)} r(t,n)
\label{eq.AAD}
\end{equation}

The first step ($\Gamma$) projects the audio signal onto a meaningful auditory representation $s(t)$ (see Fig. 1), such as the signal envelope or spectrogram (or mel-spectrogram). These two representations have meaningful links to the patterns that invoke strong responses in cortical networks in the brain that are typically observed in surface electrodes \cite{o2015attentional,crosse2016multivariate}. The second step learns a mapping $\Psi$ between the representation $s(t)$ and neural response $r(t,n)$. The linear AAD scheme employs a regression framework that minimizes the Mean Squared Error (MSE) loss between the ground truth and predicted signals. Alternative models have considered end-to-end schemes to learn nonlinear transformations using CNNs, LSTMs and self-attention networks, therefore accounting for complex relationships and variable interactions between the sensory and brain signals \cite{CNN-AAD, accou2023decoding, lu2021auditory}. Though leading to improved performance over the linear formulations, these end-to-end systems have limited generalizability due to data requirements, computational resources and need for extensive tuning and regularization. An alternative method that has been recently considered is to leverage deep methods for the initial mapping $\Gamma(.)$ ending with a linear layer for the second mapping $\Psi(.)$. This approach enables adoption of a wider range of deep methods particularly self-supervised embeddings which can be trained independently on larger datasets. This framework has been recently tested on a small scale in ECoG recordings from 3 subjects \cite{han2023improved}. The scalability of this scheme for different embedding features and larger datasets has not been evaluated before.

\begin{figure}[]
    \centering
    \includegraphics[width=1\columnwidth, trim=0 0 0 0, clip]{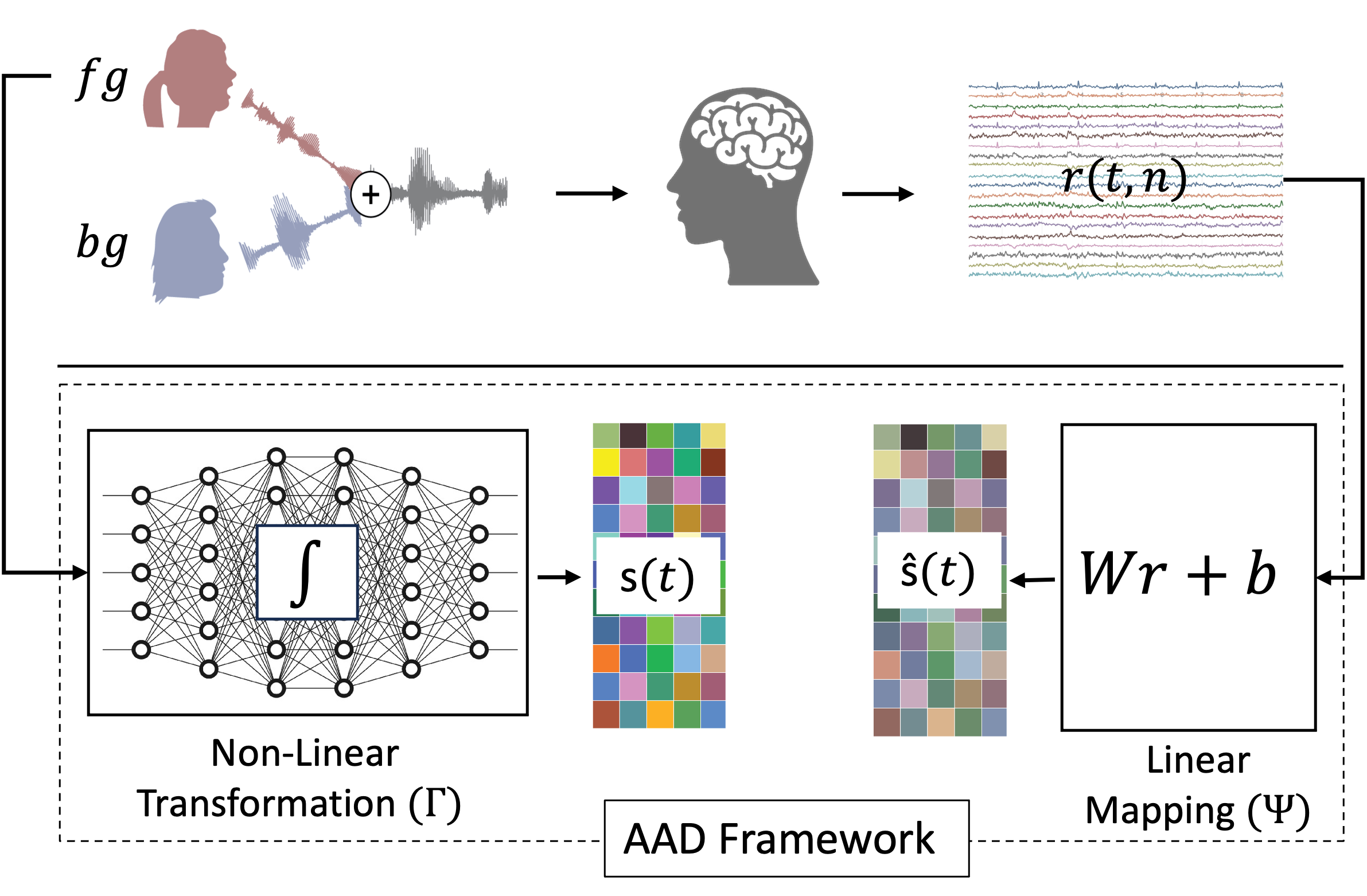}
    \caption{Auditory Attention Decoding (AAD) framework}
    \label{fig:AAD_A/U}
    \vspace{-4mm}
\end{figure}

\vspace{-.3cm}
\subsection{Meta Analysis of Sensory Representations}

The current study explores a wide range of Self Supervised (SS) representations that have recently shown promissing results in a wide range of audio and speech tasks \cite{yang2021superb}. These promise of these models is the possibility of using pre-trained transformer models that can then be leveraged for different audio and speech tasks. The current study explores the benefits of a range of these representations for AAD. Generally, SS mappings can be grouped into 2 broad categories.  Models such as ALBERT \cite{lan2019albert}, Mockingjay \cite{liu2020mockingjay}, and TERA \cite{liu2021tera} focus on reconstructing continuous filter bank features by masking parts of the input along different axes. Conversely, wav2vec2.0 \cite{baevski2020wav2vec} and HuBERT \cite{hsu2021hubert} focus on extracting discrete features from time-domain signals and learning to predict these discrete representations from masked audio signals. Notably, while wav2vec2.0 jointly trains the tasks of vector quantization and mask prediction, HuBERT employs an iterative re-clustering and re-training method for discrete representation learning and mask prediction. Based on HuBERT, a more recent development WavLM \cite{chen2022wavlm} further extends the pre-training task to both reconstruction and denoising. Table 1 summarizes all the deep representations considered for our meta-analysis. All SSL embeddings map onto a 768 dimensional representation hence allowing side-by-side comparisons of different deep features. Additionally, we also consider 'shallow' features that have been widely used in AAD tasks, notably the auditory envelope and mel-spectrogram. 


\begin{table}[!tb]
\centering
\label{table:ss_feat}
\caption{Self-Supervised Model Specifications}
\begin{tabular}{|c|c|c|c|}
\hline
\textbf{Model} & \textbf{Quantized} & \textbf{Stride} & \textbf{\# Layers} \\
\hline
AlBERT & x & 10ms & 4 \\
Mockingjay & x & 10ms & 4 \\
TERA & x & 10ms & 4 \\
\hline
HuBERT & \checkmark & 20ms & 13 \\
Wav2Vec2.0 & \checkmark & 20ms & 13 \\
WavLM & \checkmark & 20ms & 13 \\
\hline
\end{tabular}
\end{table}


\vspace{-3mm}
\subsection{Decoding Evaluations}

In a two-speaker scenario, AAD can be approached from two distinct angles, as suggested by \cite{Fuglsang2017NoiserobustCT}: attended decoding and unattended decoding. Attended decoding focuses on identifying the speaker to whom attention is directed, while unattended decoding aims to ascertain information about the unattended speaker (see Fig. \ref{fig:AAD_A/U}). In attended decoding, a trial is considered correctly decoded if the correlation between foreground and predicted foreground is greater than or equal to the correlation between background, and predicted foreground; and the opposite in unattended decoding. Studies suggest that both the attended and unattended speech streams are present in brain signals \cite{Puvvada2017CorticalRO}, further emphasizing the significance of these different decoding approaches. 


\section{Experimental Setup}
\label{sec:pagestyle}

\subsection{Neural Datasets and Preprocessing}
\label{ssec:dataset}

The evaluation covers three unique datasets each with a different language: the Fuglsang dataset (FU\_18) denoted as the DTU dataset \cite{fugslang_2018}; the Etard dataset (ET\_22) \cite{etard_octave_2022_7778289}; and the Zhang dataset (ZH\_22) \cite{zhang_yuanming_2022_7253438}, as outlined in Table 2. The trials corresponding to the repetition and single-speaker conditions were discarded from all datasets to avoid any potential leaks in the test set. To ensure a fair comparison across these datasets, we employed a standardized preprocessing pipeline using the MNE python toolbox \cite{GramfortEtAl2013a}. The standardized preprocessing involved the following steps: line noise removal, band pass filtering 0-8 Hz, artifact removal, and re-referencing based on average. All EEG channels were included in the modeling and downsampled to 64 Hz for AAD. All auditory stimuli were downsampled to 16KHz before applying different transformations for feature extraction. 

\begin{table}[]
\label{dataset_info}
\caption{Dataset Information}
\centering
\adjustbox{max width=\textwidth}{
\begin{tabular}{|c|c|c|c|c|}
\hline
\textbf{Dataset} & \textbf{\#Subjects} & \textbf{Duration} & \textbf{Language} \\
\hline
FU\_18 \cite{fugslang_2018} & 18 & 15 hrs & Dutch \\
ET\_22 \cite{etard_octave_2022_7778289} & 18 & 6 hrs & English \\
ZH\_20 \cite{zhang_yuanming_2022_7253438} & 21 & 14 hrs & Mandrian \\
\hline
\textbf{Total} & 57 & 35 hrs & - \\
\hline
\end{tabular}}
\end{table}

\subsection{Deep Feature Extraction and Preprocessing}



Each deep feature was extracted using the standard S3PRL \cite{yang2021superb} upstream configuration. Amongst different versions of the models available online, we only selected the model checkpoints that were trained on the LibriSpeech \cite{panayotov2015librispeech} corpus. We analyze two different layer combinations for each SSL model: the Last Layer (LL) and the First-Middle-Last (FML) concatenation representation. Analyzing the FML outputs allowed us to assess whether using information from multiple layers could improve AAD performance. To reduce the high dimensional embedding space of deep representations (768 dimensions), we reduced the embedding space of each layer in a model down to its 20 principal components. Leading to 20 channels for LL and 60 channels for FML configuration after concatenation. Due to different stride values (see Table 1), all features were resampled to 64 Hz. Lastly, all features underwent normalization before training.


\subsection{Shallow Feature Extraction and Preprocessing}

The speech envelope is estimated using a gamma tone filter bank of 28 filters, covering frequencies from 50 Hz to 5 kHz to capture key auditory features (Bisman et al. \cite{envelope_bisman}). To enhance the envelope first, the absolute value of each filter sample is computed to focus on signal magnitude. Then, these absolute values are exponentiated with a power factor of 0.6 to model the relationship between perceived loudness and intensity. Finally, the algorithm calculates the mean across all 28 filters to obtain the speech stimulus envelope. A 20-bin mel spectrogram is extracted using Librosa's built-in function with a 25 ms window. The choice of 20 bins is to match it along the reduced 20-dimensional space of deep representations. Both the features are downsampled to 64 Hz and normalized for AAD.

\begin{figure}[]
    \centering
    \includegraphics[scale = 0.55, trim=5 5 5 5, clip]{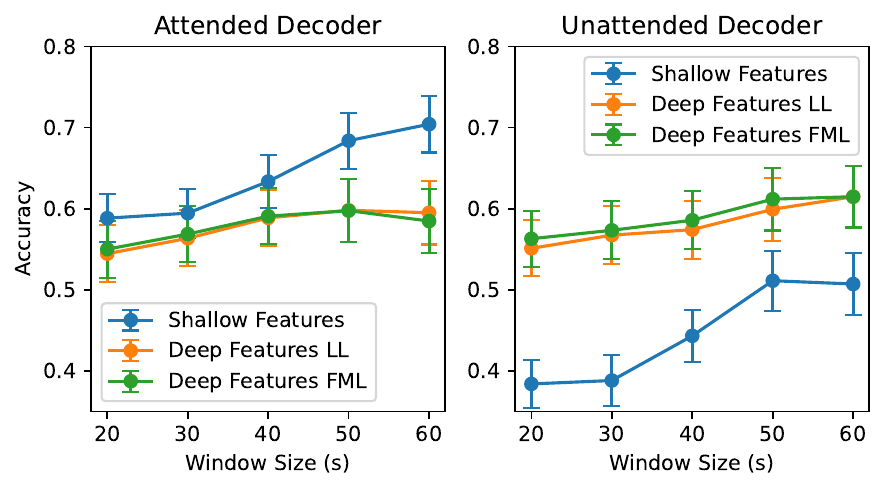}
    \caption{AAD performance of Attended and Unattended decoders with varying window sizes averaged individually for three groups across all the features.}
    \label{fig:AAD_window}
\end{figure}

\subsection{Model Training and Cross Validation}

To train the linear decoder, we harnessed the regularized TimeDelayingRidge toolbox available in MNE Python \cite{GramfortEtAl2013a}. The multivariate approach employed in this article shares similarities with the mTRF method outlined in the \cite{crosse2016multivariate}. The primary objective is to establish a linear relationship between the EEG data and the target audio features over an integration window. Ridge regression introduces a regularization term to the standard mean squared error, controlled by the hyperparameter $\lambda$. The final solution is obtained by minimizing the MSE loss between the reconstructed feature and the target feature with a given hyperparameter $\lambda$. We employed models for each subject, implementing a 90-10 train-test split for each subject's data. For hyperparameter tuning, we conducted cross-validation on each subject, utilizing leave-one-out cross-validation on the training dataset to determine the optimal $\lambda$. We choose a time-delayed window of 500ms for model training similar to \cite{Fuglsang2017NoiserobustCT}.


\begin{table*}[!htbp]
\caption{Performance comparison of Attended and Unattended Decoders across varying datasets and features.}
\label{table_example}
\centering
\renewcommand{\arraystretch}{0.95}
\adjustbox{max width=\textwidth}{
\begin{tabular}{|c|c|c|c|c|c|c|c|c|}
\hline
\multirow{2}{*}{\textbf{Feature}} & \multicolumn{4}{c|}{\textbf{Attended Decoder}} & \multicolumn{4}{c|}{\textbf{Unattended Decoder}} \\
\cline{2-9}
& \textbf{FU\_18} & \textbf{ET\_22} & \textbf{ZH\_20} & \textbf{Avg} & \textbf{FU\_18} & \textbf{ET\_22} & \textbf{ZH\_20} & \textbf{Avg} \\
\hline
Envelope & $0.65 \pm 0.36$ & \cellcolor{black!12}$0.94 \pm 0.23$ & $0.55 \pm 0.25$ & $0.69 \pm 0.34$ & $0.48 \pm 0.34$ & $0.55 \pm 0.51$ & $0.48 \pm 0.25$ & $0.49 \pm 0.36$ \\
Spectrogram & $0.70 \pm 0.32$ & $0.89 \pm 0.32$ & \cellcolor{black!12}$0.60 \pm 0.27$ & \cellcolor{black!12}\bm{$0.71 \pm 0.33$} & $0.53 \pm 0.36$ & $0.56 \pm 0.51$ & $0.46 \pm 0.26$ & $0.51 \pm 0.37$ \\
\hline
\hline
Albert\_ll & $0.73 \pm 0.30$ & $0.50 \pm 0.51$ & $0.58 \pm 0.27$ & $0.65 \pm 0.35 $ & $0.69 \pm 0.34$ & $0.61 \pm 0.50$ & $0.57 \pm 0.21$ & $0.64 \pm 0.35$ \\
Mockingjay\_ll & $0.74 \pm 0.35$ & $0.44 \pm 0.51$ & $0.46 \pm 0.30$ & $0.62 \pm 0.39$ & $0.69 \pm 0.36$ & $0.50 \pm 0.51$ & $0.54 \pm 0.27$ & $0.62 \pm 0.38$ \\
Tera\_ll & \cellcolor{black!12}$0.77 \pm 0.32$ & $0.33 \pm 0.49$ & $0.48 \pm 0.26$ & $0.62 \pm 0.38$ & \cellcolor{black!12}$0.74 \pm 0.30$ & \cellcolor{black!12}$0.78 \pm 0.43$ & $0.59 \pm 0.23$ & \cellcolor{black!12}\bm{$0.71 \pm 0.32$} \\
\hline
Hubert\_ll & $0.72 \pm 0.30$ & $0.33 \pm 0.49$ & $0.45 \pm 0.27$ & $0.58 \pm 0.37$ & $0.60 \pm 0.34$ & $0.50 \pm 0.51$ & $0.56 \pm 0.24$ & $0.57 \pm 0.36$ \\
Wav2Vec2\_ll & $0.61 \pm 0.35$ & $0.50 \pm 0.51$ & $0.47 \pm 0.24$ & $0.55 \pm 0.36$ & $0.63 \pm 0.38$ & $0.50 \pm 0.51$ & $0.59 \pm 0.30$ & $0.59 \pm 0.37$ \\
WavLM\_ll & $0.57 \pm 0.34$ & $0.50 \pm 0.51$ & $0.43 \pm 0.28$ & $0.52 \pm 0.37$ & $0.61 \pm 0.36$ & $0.33 \pm 0.49$ & $0.53 \pm 0.29$ & $0.53 \pm 0.38$ \\

\hline
\hline

Albert\_fml & $0.68 \pm 0.34$ & $0.39 \pm 0.50$ & $0.50 \pm 0.21$ & $0.58 \pm 0.37$ & $0.69 \pm 0.35$ & $0.67 \pm 0.49$ & $0.59 \pm 0.27$ & $0.66 \pm 0.36$ \\
Mockingjay\_fml & $0.68 \pm 0.39$ & $0.39 \pm 0.50$ & $0.55 \pm 0.25$ & $0.59 \pm 0.40$ & $0.62 \pm 0.39$ & $0.50 \pm 0.51$ & $0.54 \pm 0.31$ & $0.57 \pm 0.40$ \\
Tera\_fml & $0.68 \pm 0.35$ & $0.39 \pm 0.50$ & $0.55 \pm 0.28$ & $0.59 \pm 0.38$ & $0.73 \pm 0.33$ & $0.67 \pm 0.49$ & \cellcolor{black!12}$0.60 \pm 0.27$ & $0.69 \pm 0.34$ \\
\hline
Hubert\_fml & $0.70 \pm 0.33$ & $0.44 \pm 0.51$ & $0.49 \pm 0.26$ & $0.60 \pm 0.37$ & $0.64 \pm 0.37$ & $0.44 \pm 0.51$ & $0.55 \pm 0.24$ & $0.58 \pm 0.38$ \\
Wav2Vec2\_fml & $0.67 \pm 0.32$ & $0.39 \pm 0.50$ & $0.49 \pm 0.20$ & $0.57 \pm 0.35$ & $0.64 \pm 0.36$ & $0.50 \pm 0.51$ & $0.59 \pm 0.26$ & $0.60 \pm 0.37$ \\
WavLM\_fml & $0.65 \pm 0.35$ & $0.39 \pm 0.50$ & $0.48 \pm 0.25$ & $0.56 \pm 0.37$ & $0.64 \pm 0.36$ & $0.33 \pm 0.49$ & $0.63 \pm 0.26$ & $0.57 \pm 0.38$ \\
\hline
\end{tabular}
}
\end{table*}

\section{Experimental Results}

\subsection{Main Results}
\label{ssec:mainresults}

AAD performance was measured using accuracy on the subject's test trials over non-overlapping windows. In Table 3, our results across various experiments and datasets demonstrate that shallow features are better when decoding attention using the attended decoder. However, when using the unattended decoder the deep representations are better. Notably, the TERA \cite{liu2021tera} model consistently outperforms others across all datasets. This suggests that deep features have an advantage in capturing and decoding unattended signals in the brain, implying potential nonlinear encoding of auditory information in EEG similar to that of the deep features. Nonetheless, the TERA features stand out as best performer for the FU\_18 dataset. Despite deep features being pre-trained on English data, our features exhibit no bias towards English language datasets, indicating their adaptability for cross-linguistic analysis in brain signal decoding. We do not observe any consistent improvements in concatenating representations from the layers of the model across features when using linear decoders. Overall, there is no one feature that works best across the attended and unattended decoders using linear decoders on EEG data.\\

\vspace{-8mm}
\subsection{Effect of Window Size}

To further explore the factors influencing AAD performance, we investigated the impact of different window sizes on the decoding accuracy. We varied the window sizes while keeping the representation type constant and analyzed the resulting performance differences. Our analysis revealed that window size had a significant influence on AAD performance. Specifically, larger window sizes tended to yield higher decoding accuracy, suggesting that a broader temporal context enhances the ability to isolate desired sound sources (see Fig. \ref{fig:AAD_window}). In addition, we also observe that the linear decoding accuracy of the unattended decoder for deep features is consistently better across all window sizes. And there exists no significant improvement in adding information from different layers of the model across different window sizes.

\begin{figure}[!tb]
    \centering
    \includegraphics[width=.9\columnwidth, trim=30 20 60 150, clip, scale = 3]{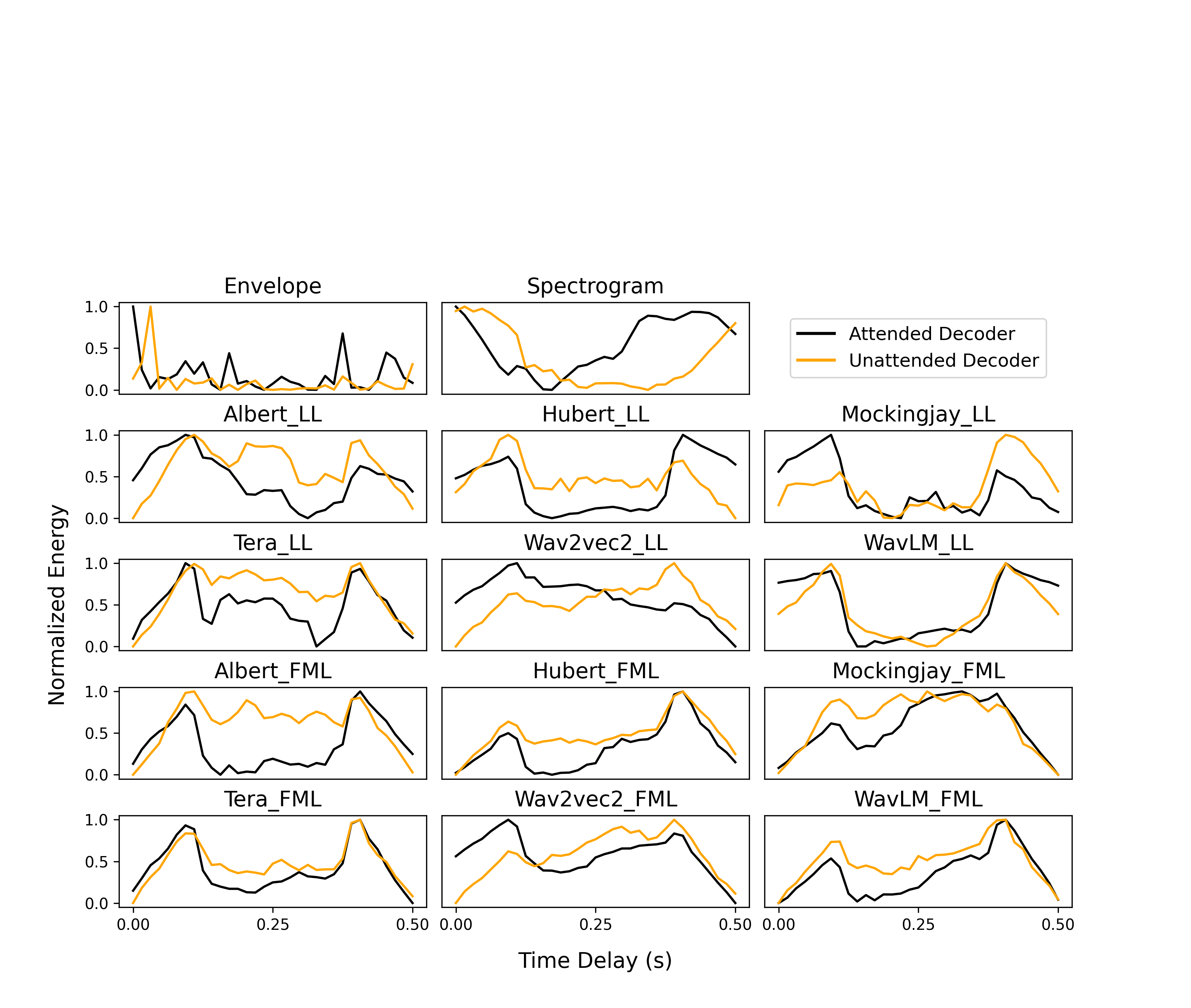}
    \caption{Normalized energy plots of model weights when averaged across channels.}
    \label{fig:model_weight}
\end{figure}

\vspace{-4mm}
\subsection{Investigating the Model Weights}

Examining the normalized energy of linear model weights across various time delays (0-500ms) in our study revealed insights into the temporal dynamics of AAD. The energy distribution of initial time delays exhibited higher weighting indicating a marker for auditory attention at the start. However, the weightage decreases from 200 to 300 ms and then peaks higher indicating the discrimination between attended and unattended is maximal at a time delay beyond 300ms time delay. This observation is dominant across many features as observed in Fig. \ref{fig:model_weight}.

\section{Conclusion}

The study offers a meta-analysis of different mapping of audio stimuli on large EEG data. By leveraging a final linear layer, the analysis provides a controlled comparison across shallow and deep embeddings and reveals the possible value of nonlinear mappings in unraveling different mechanisms of encoding foreground and background information in the brain. As this technology continues to evolve, these findings open exciting possibilities for the exploration of new features learned by large deep neural networks for improving AAD.







\small
\bibliographystyle{IEEEbib}
\bibliography{refs}

\end{document}